\documentclass[aps,twocolumn,showpacs,amssymb,pra]{revtex4-1}
\usepackage{amsmath}
\usepackage{hyperref}
\usepackage{graphicx}
\usepackage{bbm}
\usepackage{color}

\usepackage{bm}

\usepackage[normalem]{ulem}

\allowdisplaybreaks

\begin{document}
\title{Dynamic response of spin-2 bosons in one-dimensional optical lattices}

\author{Florian Lange}
\affiliation{Institut f\"ur Physik, Universit\"at
Greifswald, 17489 Greifswald, Germany}
\author{Satoshi Ejima}
\affiliation{Institut f\"ur Physik, Universit\"at
Greifswald, 17489 Greifswald, Germany}
\author{Holger Fehske}
\affiliation{Institut f\"ur Physik, Universit\"at
Greifswald, 17489 Greifswald, Germany}

\begin{abstract}
  We investigate the spin-2 chain model corresponding to the small hopping limit of the spin-2 Bose-Hubbard model using density-matrix renormalization-group and time-evolution techniques. We calculate both static correlation functions and the dynamic structure factor. 
  The dynamic structure factor in the dimerized phase differs significantly between parameters near the SU(5)-symmetric point and those deeper in the phase where the dimerization is strong. 
  In the former case, most of the spectral weight is concentrated in a single excitation line, while in the latter case, a broad excitation continuum shows up. 
  For the trimerized phase, we find gapless excitations at momenta $k=\pm2\pi/3$ in agreement with previous results, although the visibility of these excitations in the dynamic spin response depends strongly on the specific parameters. 
 We also consider parameters for specific atoms which may be relevant for future optical-lattice experiments. 
\end{abstract}

\maketitle

\section{Introduction}
After the realization of the Bose-Hubbard model and its superfluid-Mott insulator transition~\cite{Greiner2002}, there have been many proposals to extend experiments with optical lattices
to other systems~\cite{Gross995,Simon2011}. 
One approach is to make use of the hyperfine spin of alkali-metal atoms to add a spin-1 or spin-2 degree of freedom to the particles~\cite{PhysRevLett.88.163001,SpinorBoseGasReview}, as has already been done in experiments with Bose-Einstein condensates~\cite{Stenger1998,PhysRevLett.92.040402,PhysRevLett.92.140403,PhysRevA.69.063604}. 
Such systems are expected to be described by generalizations of the Bose-Hubbard model with additional spin-dependent interactions. 
These interactions could give rise to much richer phase diagrams, which makes the models interesting also from a theoretical point of view~\cite{PhysRevB.77.014503,SpinorBosonsClassification,Spin2BoseHubbardRigorous}. 

The Mott insulating phases in a deep optical lattice can be studied more easily in effective models of localized spins~\cite{PhysRevA.68.063602}. 
Here we are interested in the spin-2 chain corresponding to spin-2 bosons in a one-dimensional lattice at unit filling. 
In a mean-field approximation this model realizes ferromagnetic, nematic, and cyclic phases that each break the spin-rotation symmetry in a different way~\cite{SpinorBosonsClassification,Spin2BoseHubbardNJP}. 
However, a more reliable density-matrix renormalization-group (DMRG) study showed that
in one dimension the nematic and cyclic phases are replaced, respectively, by dimerized and trimerized phases conserving the spin-rotation symmetry~\cite{Spin2BoseHubbardDMRG1}. 
This is in agreement with the Mermin-Wagner theorem which forbids the spontaneous breaking of the continuous spin-rotation symmetry in the case of nematic or cyclic order.

While the phase diagram has been established, 
the static and dynamic properties of the spin-2 chain are much less explored, also in comparison with its spin-1 counterpart. 
In particular, the dynamic response should be of interest in the case where the model could be realized experimentally. 
For this reason, the primary objective of this paper is to calculate the dynamic spin structure factor, which gives valuable insight into the excitation spectrum of this system, and should be accessible in future experiments~\cite{SpinDependentBragg}.  We restrict ourselves to the dimerized and trimerized phases specific to one dimension.

\section{Model and method}
Bosonic atoms with a fixed hyperfine spin $S=2$ in an optical lattice are expected to be described by the spin-2 Bose-Hubbard model
\begin{align}
\hat{H}_B &= -t \sum_{j \sigma} (\hat{b}_{j \sigma}^{\dagger} \hat{b}_{j+1, \sigma}^{\phantom{\dagger}} + {\rm H. c.}) + \sum_j \sum_{n=0,2,4} g_n \hat{P}_{j}^n \, , 
\end{align}
where $\hat{b}_{j \sigma}^{\dagger}$ $(\hat{b}_{j \sigma}^{\phantom{\dagger}})$ are bosonic creation (annihilation) operators and $\sigma \in \{-2,-1,0,1,2\}$ is the $z$ projection of the hyperfine spin. The interaction term consists of projection operators $\hat{P}_{j}^n$ onto the subspace of states with total spin $n$ at site $j$. 
It describes $s$-wave scattering between the particles, and the interaction strengths $g_n$ are, up to constant factor, the scattering lengths for the corresponding channel~\cite{Spin2BoseHubbardNJP}. 

\begin{figure}[tb]
\includegraphics[width=0.75\linewidth]{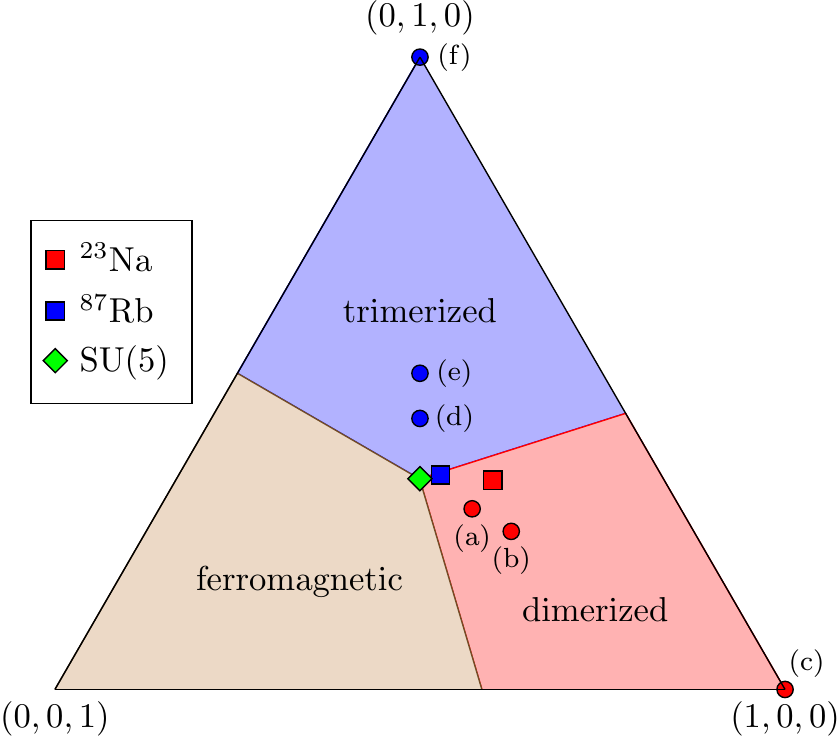}
\caption{
Schematic phase diagram of the model~\eqref{eqham} as a ternary plot of the variables $(\epsilon_0,\epsilon_2,\epsilon_4)/(\epsilon_0+\epsilon_2+\epsilon_4)$~\cite{Spin2BoseHubbardDMRG1}. 
The circles labeled (a)-(f) indicate the model-parameter values used in the corresponding panels of Fig.~\ref{figSkw}. }
\label{figPD}
\end{figure}

We study the limit of small hopping at unit filling and 
assume that the interaction strengths are such that the ground state has a uniform density. 
The effective spin-2 chain for this limit in second-order perturbation theory is  
\begin{align}
\hat{H} &= \sum_j \sum_{n=0,2,4} \epsilon_n \hat{P}_{j,j+1}^n \, .
\label{eqham}
\end{align}
Here, $\hat{P}_{j,j+1}^n$ is the projection operator onto the subspace of states with total spin $n$ between sites $j$ and $j+1$, and $\epsilon_n = -4t^2/g_n$.  
We assume $\epsilon_n < 0$ since other parameter regions are not accessible with spin-2 bosons. 
The phase diagram for this model obtained in Ref.~\cite{Spin2BoseHubbardDMRG1} can be summarized as follows (cf. Fig.~\ref{figPD}): 
If the term proportional to $\epsilon_0$ is dominant, the system is in a spontaneously dimerized gapped phase. The $\epsilon_2$ term instead favors a gapless phase which has a trimerized ground state for finite systems. Finally, a sufficiently large $\epsilon_4$ term leads to ferromagnetic order.

The dimerized and trimerized phases both have the full spin-rotation symmetry in the ground state. 
In the dimerized phase, the symmetry under translation by one site is spontaneously broken while the symmetry under bond-centered reflection is conserved. 
This is captured by the order parameter 
\begin{align}
{\cal O}_D &= |\langle \hat{h}_j - \hat{h}_{j+1} \rangle| / |\langle \hat{h}_j + \hat{h}_{j+1} \rangle| \, ,
\label{eqOP}
\end{align}
where $\hat{h}_j=\sum_{n=0,2,4}\epsilon_n \hat{P}_{j,j+1}^n$ is the nearest-neighbor term in the Hamiltonian acting on the sites $j$ and $j+1$.  
A dimerized phase also occurs in the model describing spin-1 bosons, the spin-1 bilinear-biquadratic chain~\cite{PhysRevB.40.4621}. 
It has long been debated for this model whether there is a direct transition to the ferromagnet or an intermediate disordered nematic phase exists~\cite{PhysRevB.43.3337,BilinearBiquadraticAbsenceNematic,Spin1BBQuadrupolarCorrelations}. 
Recent numerical calculations indicate the absence of a nematic phase but find a very small dimerization near the transition~\cite{Spin1DimerBerryPhase}. 
Here, we take a similar view for the spin-2 model, although the distinction between a weakly dimerized phase and a uniform nematic phase is difficult to detect numerically. 

The name of the trimerized phase originates from the period-3 structure seen for finite systems in the bond observables such as nearest-neighbor spin correlations~\cite{Spin2BoseHubbardDMRG1}. In the thermodynamic limit, this structure disappears and the lattice symmetry is unbroken. Additionally the excitation gap vanishes unlike in the dimerized phase. 
The trimerized phase does not occur in spin-1 bosons but resembles the gapless phase in a different parameter region of the bilinear-biquadratic chain. 
It was shown numerically to be described by the SU(3)${}_1$ Wess-Zumino-Witten field theory with central charge $c=2$~\cite{Spin2BoseHubbardDMRG2}. 
In the same work, exact-diagonalization spectra were provided which exhibit minima at $k=\pm 2\pi/3$. The excitations at these momenta are expected to become gapless in the thermodynamic limit, which can serve as a signature of the phase in the dynamic spin response. 

At the point $\epsilon_0=\epsilon_2=\epsilon_4$, where the three phases meet, the symmetry of the Hamiltonian \eqref{eqham} becomes SU(5) and the ground state is highly degenerate~\cite{SpinorBosonsClassification,Spin2BoseHubbardRigorous}. 
The degeneracy is lifted, however, when moving into the dimerized or the trimerized phase. 
Only a twofold degeneracy due to the broken translation symmetry remains in the dimerized phase. 
In the spinful Bose-Hubbard model, from which the effective Hamiltonian~\eqref{eqham} is derived, the SU(5)-symmetric point corresponds to the absence of any spin-dependent interactions. 

Accurate numerical results for the ground states of one-dimensional systems can be obtained with the DMRG which is based on a matrix-product-state (MPS) ansatz~\cite{White92,Sch11}. 
Here, we employ the infinite DMRG (iDMRG) that 
works directly in the thermodynamic limit and approximates the ground state by an infinite MPS (iMPS)~\cite{Mc08,iTEBD}. 
Similarly to the finite-system DMRG, the accuracy of the approximation is determined by the so-called bond-dimension $\chi$. For details of the numerical method, see Ref.~\cite{Sch11}. 
The iMPS ansatz is well suited to describe gapped ground states but cannot capture the power-law decay of correlations in critical phases. 
Nevertheless, even for gapless states the correlation functions are correctly reproduced up to a finite distance that increases with the bond dimension $\chi$~\cite{iTEBD}. 
It is therefore possible to obtain reliable information about the critical properties with the iDMRG method~\cite{10.21468/SciPostPhys.5.6.059}. 

\begin{figure}[tb]
\includegraphics[width=0.8\linewidth]{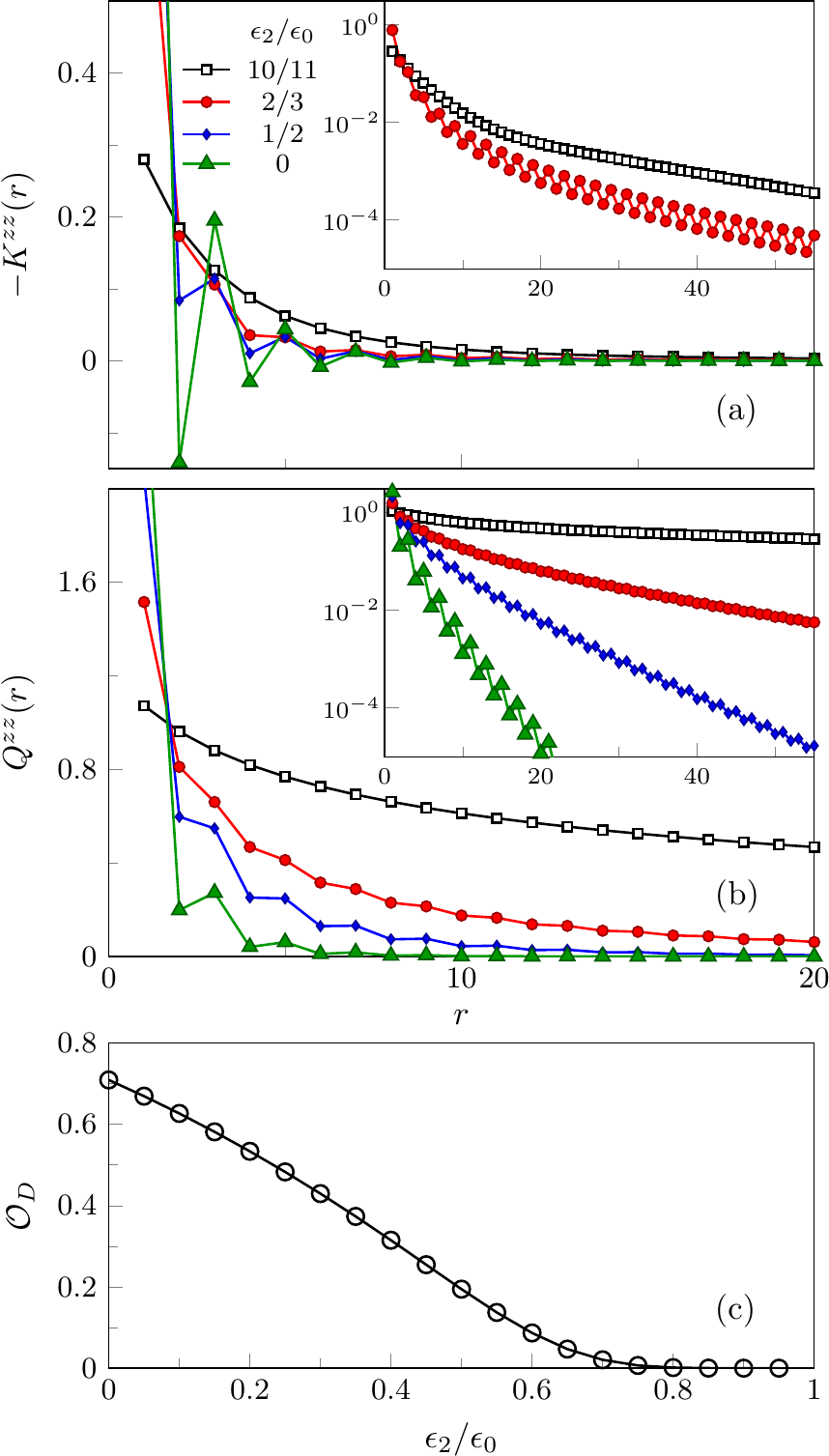}
\caption{
(a) Spin-spin and (b) quadrupolar correlation functions defined in Eq.~\eqref{eqCFK} and~\eqref{eqCFQ} for the dimerized phase. A bond dimension of $\chi=4000$ was used in the iDMRG calculations. 
The insets display the same results using a semi-logarithmic representation. (c) Dimerization order parameter ${\cal O}_D$~\eqref{eqOP} as a function of $\epsilon_2/\epsilon_0$.  }
\label{figCF_D}
\end{figure}

Static correlation functions can be calculated directly from the iMPS ground state. 
To obtain the dynamic structure factors, we use the iMPS as input for a time-evolving-block-decimation simulation~\cite{PhysRevLett.91.147902} with infinite boundary conditions~\cite{InfiniteBoundary1}. 
We spread the time evolution to two separate states in order to reach longer times and thereby a better resolution in frequency space~\cite{TimeEvolutionTwoStates,PhysRevB.97.060403}. 
Furthermore, we use linear prediction to extrapolate the calculated dynamic correlation functions to longer times~\cite{PhysRevB.77.134437}. This can be done reliably, if the spectrum consists of a small number of sharp excitation peaks. \\

\section{Static correlations}
Figure~\ref{figCF_D} shows the iDMRG results for the static spin-spin correlation function
\begin{align}
K^{zz}(r) &= \langle \hat{S}_{j+r}^z \hat{S}_j^z \rangle \, ,
\label{eqCFK}
\end{align}
and for the quadrupolar correlation function
\begin{align}
Q^{zz}(r) &= \langle [(\hat{S}_{j+r}^{z})^2-2] [(\hat{S}_{j}^{z})^2-2] \rangle \, ,
\label{eqCFQ}
\end{align}
in the dimerized phase.  
The latter is of interest, since quadrupolar ordering occurs in the nematic phase for similar parameters in higher-dimensional versions of the model~\cite{SpinorBosonsClassification}. 
For simplicity, we consider only parameter points on the line $\epsilon_2=\epsilon_4$. 
Since the dimerized phase is gapped, the correlations fall off exponentially at long distances. Near the SU(5) point $\epsilon_2/\epsilon_0=1$, however, the correlation length is quite large, as can be seen in the quadrupolar correlations $Q^{zz}(r)$. 
Both functions $K^{zz}(r)$ and $Q^{zz}(r)$ are more or less smooth for $\epsilon_2/\epsilon_0 \lesssim 1$ but develop a period-2 structure when $\epsilon_2/\epsilon_0$ is decreased. 
This is indicative of the dimerization, which can be more clearly detected by the order parameter ${\cal O}_D$ defined in Eq.~\eqref{eqOP} [see Fig.~\ref{figCF_D}(c)]. 
We find that ${\cal O}_D$ is almost zero for $\epsilon_2/\epsilon_0 \gtrsim 0.7$ but quickly increases for smaller values. Similar behavior of the order parameter ${\cal O}_D$ and dominance of quadrupolar correlations have also been observed in the spin-1 bilinear-biquadratic chain near the transition between ferromagnet and dimerized phases~\cite{Spin1DimerBerryPhase,Spin1BBQuadrupolarCorrelations}. 

\begin{figure}[tb]
\includegraphics[width=0.8\linewidth]{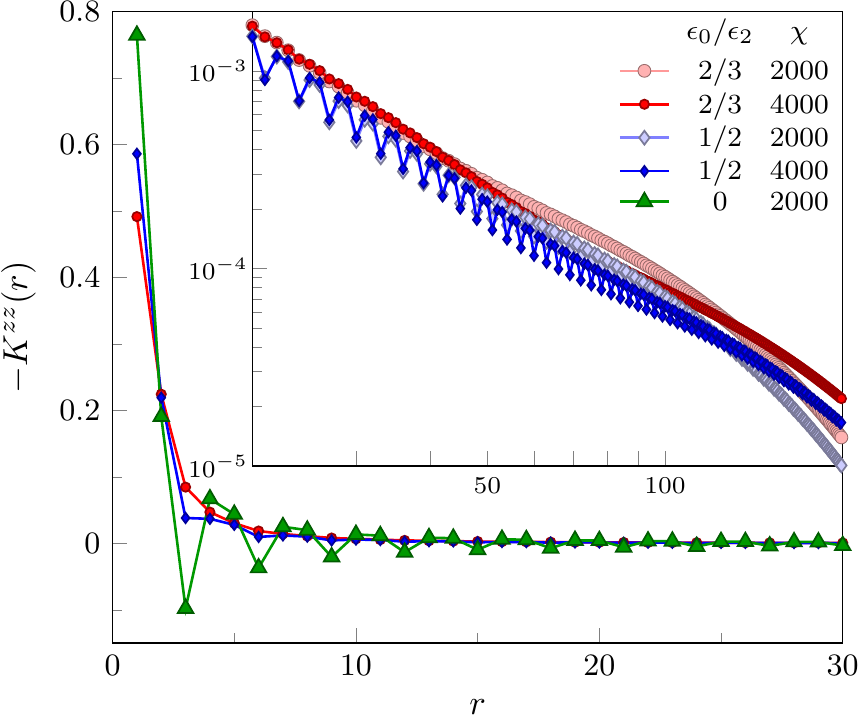}
\caption{Spin-spin correlation function~\eqref{eqCFK} in the trimerized phase. The inset uses a log-log scale for the same data. }
\label{figCF_T}
\end{figure}

Results for the trimerized phase are displayed in Fig.~\ref{figCF_T}. Here, we choose $\epsilon_0=\epsilon_4$ and analyze the dependence on $0 \leq \epsilon_0/\epsilon_2 < 1$. 
The spin-spin correlations again fall off smoothly near the SU(5) point but now show a period-3 structure deeper in the phase. 
In contrast to the dimerized phase, the correlations decrease with a power-law, as can be seen in the inset of Fig.~\ref{figCF_T}. Note that the quadrupolar correlation functions do not decrease noticeably slower than the spin-spin correlations (not shown). 

The DMRG implementation explicitly enforces the U(1) symmetry of $S^z$-conservation but not the full SU(2) symmetry of spin rotations. 
Nevertheless, the ground state approximation fulfills the spin-rotation symmetry to high accuracy in the dimerized phase. In the gapless trimerized phase, the iDMRG converges to a state with broken spin symmetry. 
However, the dipolar and quadrupolar order parameters vanish, i.e., $\langle \hat{S}_j^{\alpha} \rangle = 0$ and $\langle \hat{Q}_j^{\alpha,\beta} \rangle = 0$, where 
$\hat{Q}_j^{\alpha,\beta} = \hat{S}_j^{\alpha}\hat{S}_j^{\beta} + \hat{S}_j^{\beta}\hat{S}_j^{\alpha} - 4$. 
The symmetry breaking shows up only in higher powers of the spin operators, e.g. $\langle (\hat{S}_j^z)^3 \rangle \neq \langle (\hat{S}_j^x)^3 \rangle$.
This is likely related to the fact that the trimerized phase replaces the cyclic phase in higher dimensions, where the spin-rotation symmetry breaks without dipolar and quadrupolar order occuring~\cite{PhysRevLett.97.180411,SpinorBosonsClassification}. 
These discrepancies become smaller with increasing bond dimension $\chi$, and are expected to vanish for $\chi \rightarrow \infty$. Since we are mainly interested in the dynamic spin-spin correlations, the artificial symmetry breaking should not be problematic.

\section{Dynamic spin structure factor}
The dynamic spin structure factor for a periodic chain with $N$ sites is defined by
\begin{align}
S(k,\omega) &= \sum_{n\neq0} | \langle n |  \tilde{S}_k^z | 0 \rangle |^2 \delta(\omega - (E_n-E_0))\, ,
\label{eqSkw}
\end{align}
where $\tilde{S}_k^z = (1/\sqrt{N})\sum_j e^{ikj} \hat{S}_j^z$, and $E_0$ ($E_n$) is the energy of the ground state ($n$th excited state). 
Since the Hamiltonian conserves the spin-rotation symmetry, it is not necessary to consider the other spin components separately. In our numerical calculations, we consider the thermodynamic limit $N\rightarrow \infty$. 

At the SU(5)-symmetric point $\epsilon_0=\epsilon_2=\epsilon_4$, the Hamiltonian can be written as $\hat{H}= (\epsilon_0/2) \sum_j (1 + \hat{\cal P}_{j,j+1})$, where $\hat{\cal P}_{j,j+1}$ exchanges the states of sites $j$ and $j+1$. 
We therefore have elementary excitations with dispersion
\begin{align}
\omega(k)/|\epsilon_0| &= 1-\cos(k) \, ,
\label{eqSU5}
\end{align}
which show up in $S(k,\omega)$ as $\delta$ peaks. 
In the following we will analyze how the dynamic spin response changes when moving away from this point into either the dimerized or the trimerized phase, 
again concentrating on parameters $\epsilon_2=\epsilon_4$ and $\epsilon_0=\epsilon_4$.

\begin{figure}[tb]
\includegraphics[width=0.99\linewidth]{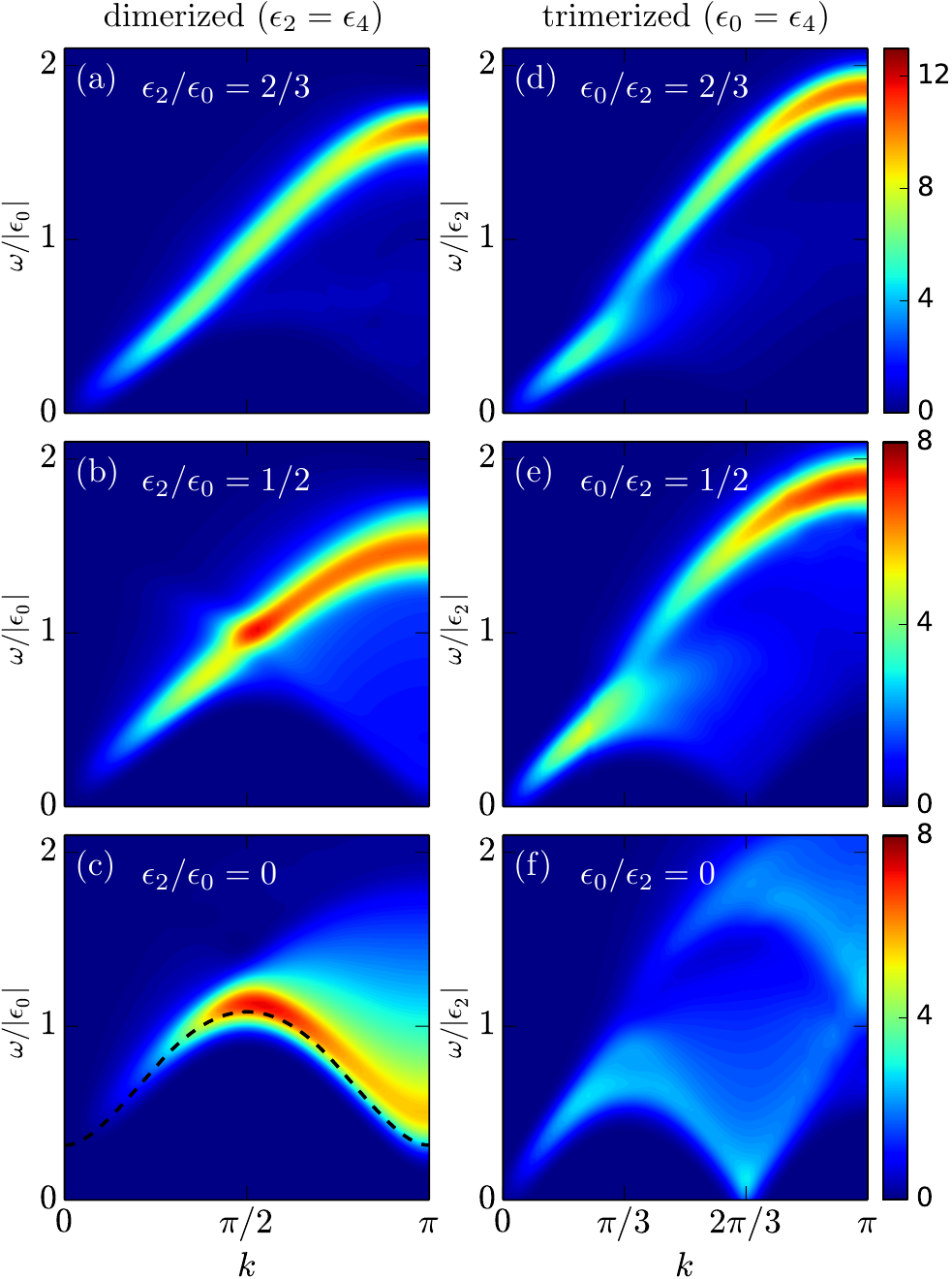}
\caption{Dynamic structure factor $S(k,\omega)$ [Eq.~\eqref{eqSkw}] in the (a)-(c) dimerized and (d)-(f) trimerized phases. The parameters used are indicated in the phase diagram of Fig.~\ref{figPD}. In (c) the exact onset of the excitation continuum according to Eq.~\eqref{eqexactdisp} is marked by the dashed line.  The energy unit is $|\epsilon_0|$ for the dimerized and $|\epsilon_2|$ for the trimerized phase. 
The spectral functions are convolved with a Gaussian function with $\sigma=0.075$ using the same energy scale. 
}
\label{figSkw}
\end{figure}

\subsection{Dimerized phase}
Let us begin by discussing the dimerized phase. 
It is reasonable to assume that the dynamic structure factor 
close to the SU(5) point shows a dispersion similar to Eq.~\eqref{eqSU5}. 
On the other hand, the excitation spectrum at the point $\epsilon_2=\epsilon_4=0$ 
is known exactly and it differs significantly from the one at the SU(5) point. In particular, it is built from pairs of excitations, 
which lead to a continuum in $S(k,\omega)$. 
Their dispersion is given by~\cite{Klumper_1990}
\begin{align}
\omega(k)/|\epsilon_0| = \sqrt{A + B \sin^2(k)} \, ,
\label{eqexactdisp}
\end{align}
where $A\approx 0.290$ and $B\approx 9.725$. 
Our iDMRG results indicate that the ground state for $\epsilon_2=0$ is strongly dimerized, nearly consisting of fully decoupled pairs of nearest-neighbor singlets. 
In fact, the exact dispersion~\eqref{eqexactdisp} roughly agrees with a simple estimate based on a decoupled site moving as a domain wall through such a fully dimerized state.

\begin{figure}[tb]
\includegraphics[width=0.99\linewidth]{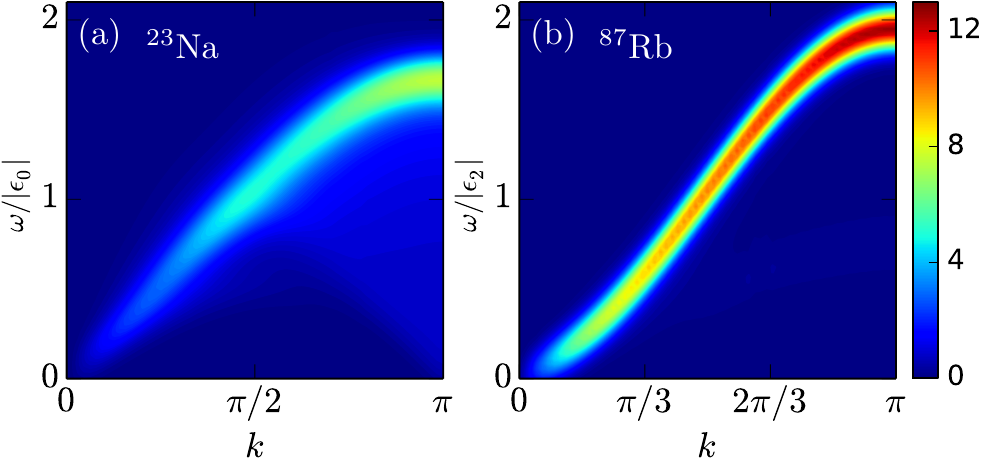}
\caption{Dynamic structure factor for parameters calculated using the scattering lengths of Ref.~\cite{Spin2BosonsScatteringLength}, namely $(\epsilon_0,\epsilon_2,\epsilon_4)/(\epsilon_0+\epsilon_2+\epsilon_4)\approx(0.43,0.33,0.24)$ and $(0.36,0.34,0.30)$ for ${}^{23}$Na and ${}^{87}$Rb, respectively. 
The Gaussian broadening is as in Fig.~\ref{figSkw}. 
}
\label{figSkwExp}
\end{figure}

Determining  numerically the dynamic structure factor $S(k,\omega)$, we can demonstrate how the excitation spectrum changes between the two limits [Figs.~\ref{figSkw}(a)-\ref{figSkw}(c)]. 
Near the SU(5) point, up to at least $\epsilon_{2}/\epsilon_0=2/3$, the dynamic structure factor $S(k,\omega)$ is indeed dominated by a single excitation line, although a broad continuum below it is also visible. 
The dispersion becomes linear at small momenta but otherwise stays qualitatively similar to the cosine form of Eq.~\eqref{eqSU5}. 
It is in fact very similar to the dispersion $\omega(k) \propto \sqrt{ [1-\cos(k)]^2 + A \sin^2(k) }$ resulting from a generalized spin-wave analysis around a nematically ordered mean-field state of a spin-1 chain~\cite{Matveev1973,PAPANICOLAOU1988367}, even though our ground state has no nematic order and we consider a spin-2 model.

Going to smaller values of $\epsilon_2/\epsilon_0$, the dynamic response changes more significantly. The energy gap becomes noticeably larger and the spectral weight gets spread over a wide excitation continuum, particularly for $|k| > \pi/2$. 
The onset of the continuum in the limit $\epsilon_2=0$ is in excellent agreement with the exact dispersion~\eqref{eqexactdisp}. 
Comparing with Fig.~\ref{figCF_D}, we find that the change in $S(k,\omega)$  
coincides with an increase in the dimerization strength ${\cal O}_D$. 

\subsection{Trimerized phase}
Let us now discuss the dynamic structure factor in the trimerized phase [Figs.~\ref{figSkw}(d)-~\ref{figSkw}(f)]. 
According to a  previous analysis, this phase is characterized by gapless excitations with spin $S=0,1,2$ at momenta $k = \pm 2\pi/3$~\cite{Spin2BoseHubbardDMRG2}. 
Numerically, we find that for $\epsilon_0/\epsilon_2=2/3$, the spectral weight is still concentrated in a single line with a dispersion similar to that found in the dimerized phase.  
Further away from the SU(5) point, for $\epsilon_0/\epsilon_2=1/2$, a continuum of  excitations appears at lower energies. 
In particular, the gap closes at $k=2\pi/3$ as anticipated in Ref.~\cite{Spin2BoseHubbardDMRG2}. 
Moving towards the limit $\epsilon_0=0$, the response at $k=2\pi/3$ becomes more pronounced. 

\subsection{Relation to experiments}
The parameters describing an optical lattice system will depend on the scattering lengths of the particles. 
With the values given in Ref.~\cite{Spin2BosonsScatteringLength}, one expects that ${}^{23}$Na atoms develop a dimerized state and ${}^{87}$Rb atoms a trimerized state~\cite{Spin2BoseHubbardDMRG1}. 
We have included results for these parameters as examples for the two phases. 
One should note, however, that 
experiments have found a very short lifetime for ${}^{23}$Na gases in the manifold with hyperfine spin $S=2$, which makes an actual realization of the corresponding spin Hamiltonian unlikely~\cite{PhysRevLett.90.090401}. 
Systems of ${}^{87}$Rb atoms are more promising~\cite{Widera_2006} although the model~\eqref{eqham} has not implemented so far. 
There are different estimates for the scattering length of ${}^{87}$Rb in the literature~\cite{SpinorBoseGasReview,Widera_2006} but the deviations are rather small and we do not expect them to notably affect the dynamic response function. 

For the ${}^{23}$Na parameters [Fig.~\ref{figSkwExp}(a)], 
the dynamic structure factor $S(k,\omega)$ seems to exhibit signatures of both the weak and the strong dimerization limit, i.e., there is a clear excitation line but also significant spectral weight in the continuum below it. 
In agreement with this, the dimerization order parameter takes an intermediate value ${\cal O}_D \approx 0.14$. 
For ${}^{87}$Rb [Fig.~\ref{figSkwExp}(b)], we do not see the low-energy excitations at $k=2\pi/3$ characteristic of the trimerized phase. Instead, the dynamic structure factor resembles that at the SU(5)-symmetric point with dispersion~\eqref{eqSU5}.  Perhaps this is not surprising since, as shown in Fig.~\ref{figPD}, 
the ${}^{87}$Rb parameters lie close to the SU(5) point.

\section{Conclusion}
We have used time-dependent matrix-product-state techniques to study the dynamic structure factor of a spin-2 chain describing spinful bosons in optical lattices. 
The spectra in the dimerized and trimerized phases are known to be qualitatively different. While the dimerized phase is gapped, the trimerized one has gapless excitations at momenta $k=\pm 2\pi/3$. 
In the dynamic spin structure factor, however, these differences become apparent only deeper into the respective phases. 
Near the SU(5) point, where dimerized, trimerized and ferromagnetic phases meet, the observed spectra are quite 
similar, with a single dominant excitation line and, in the dimerized phase, only a very small gap. 
As parameters further away from this point correspond to relatively strong spin-dependent interactions in the underlying spin-2 Bose-Hubbard model, they may be difficult to realize experimentally. 
Using the scattering lengths of Ref.~\cite{Spin2BosonsScatteringLength}, we have carried out simulations for ${}^{87}$Rb and ${}^{23}$Na. 
The dynamic structure factor for the potentially feasible ${}^{87}$Rb systems indeed shows only a single branch with a dispersion similar to the one at the SU(5) point.

So far we have considered only systems at zero temperature in the limit of a deep optical lattice. 
In a real experiment, however, temperature and hopping will be finite and it would be interesting to see how this affects the system's properties. 
While it is possible to do this with matrix-product-state techniques, the required computational effort would be significantly higher than in the present work.

\section*{Acknowledgments}
F.L. was supported by Deutsche Forschungsgemeinschaft (Germany) through Project No. FE 398/8-1. 
The DMRG simulations were performed using the ITENSOR 
library~\cite{ITensor}.

\end{document}